\documentclass{jfm}

\pdfoutput=1

\usepackage{graphicx}
\usepackage{natbib}
\usepackage{amsmath}
\usepackage{subfig}

\ifCUPmtlplainloaded \else
  \checkfont{eurm10}
  \iffontfound
    \IfFileExists{upmath.sty}
      {\typeout{^^JFound AMS Euler Roman fonts on the system,
                   using the 'upmath' package.^^J}%
       \usepackage{upmath}}
      {\typeout{^^JFound AMS Euler Roman fonts on the system, but you
                   dont seem to have the}%
       \typeout{'upmath' package installed. JFM.cls can take advantage
                 of these fonts,^^Jif you use 'upmath' package.^^J}%
      }
  \else
  \fi
\fi

\ifCUPmtlplainloaded \else
  \checkfont{msam10}
  \iffontfound
    \IfFileExists{amssymb.sty}
      {\typeout{^^JFound AMS Symbol fonts on the system, using the
                'amssymb' package.^^J}%
       \usepackage{amssymb}%
       \let\le=\leqslant  
       \let\ge=\geqslant  
      }{}
  \fi
\fi

\ifCUPmtlplainloaded \else
  \IfFileExists{amsbsy.sty}
    {\typeout{^^JFound the 'amsbsy' package on the system, using it.^^J}%
     \usepackage{amsbsy}}
    {\providecommand\boldsymbol[1]{\mbox{\boldmath $##1$}}}
\fi

\newsavebox{\astrutbox}
\sbox{\astrutbox}{\rule[-5pt]{0pt}{20pt}}

\title[Variation on the Kolmogorov Forcing]{Variation on the Kolmogorov Forcing: Asymptotic Dissipation Rate Driven by Harmonic Forcing}

\author[B. Rollin, Y. Dubief and C.R. Doering]%
{B. \ls R\ls O\ls L\ls L\ls I\ls N$^1$%
  \thanks{To whom correspondence should be addressed},\ns Y. \ls D\ls U\ls B\ls I\ls E\ls F$^1$ \and C.R. \ls  D\ls O\ls E\ls R\ls I\ls N\ls G$^2$}

\affiliation{$^1$School of Engineering, University of Vermont,
Burlington, VT 05405, USA\\[\affilskip]
$^2$Department of Mathematics, University of Michigan, Ann Arbor, MI 48109-1109, USA}

\pubyear{1996}
\volume{538}
\pagerange{119--126}
\date{?? and in revised form ??}

\begin{document}

\maketitle

\begin{abstract}
The relation between the shape of the force driving a turbulent flow and the upper bound on the dimensionless dissipation factor $\beta$ is presented. We are interested in non-trivial (more than two wave numbers) forcing functions in a three dimensional domain periodic in all directions. A comparative analysis between results given by the optimization problem and the results of Direct Numerical Simulations is performed. We report that the bound on the dissipation factor in the case of infinite Reynolds numbers have the same qualitative behavior as for the dissipation factor at finite Reynolds number. As predicted by the analysis, the dissipation factor depends strongly on the force shape. However, the optimization problem does not predict accurately the quantitative behavior. We complete our study by analyzing the mean flow profile in relation to the Stokes flow profile and the optimal multiplier profile shape for different force-shapes. We observe that in our 3D-periodic domain, the mean velocity profile and the Stokes flow profile reproduce all the characteristic features of the force-shape. The optimal multiplier proves to be linked to the intensity of the wave numbers of the forcing function. 
\end{abstract}

\section{Introduction}

In the first half of the twentieth century, authors such as \cite{richardson1922wpn}, \cite{taylor1938st} and \cite{kolmogorov1941lst} developed a description of turbulence based on the concept of an energy cascade. In this defining description of the turbulence phenomenon, kinetic energy is transferred at a constant rate from larger unstable eddies to smaller eddies. The cascade lasts until the eddy motion becomes stable and viscosity can effectively dissipate the kinetic energy. The rate of dissipation of kinetic energy $\varepsilon$ is defined by:
\begin{equation}
\varepsilon=2\nu \langle S_{ij}S_{ij}\rangle 
\end{equation}
where $S_{ij}$ is the rate of strain tensor and $\nu$ the kinematic viscosity coefficient. Analyses of the Navier-Stokes equations show that bounds on the kinetic energy dissipation rate can be derived directly from these governing equations without additional hypothesis or approximations for statistically stationary incompressible flows. During the 1990s, rigorous bounds for different types of boundary driven flows were derived in the asymptotic case of  an infinite Reynolds number $Re$ \cite[]{doering1992eds,marchioro1994red,wang1997tae}:
\begin{equation}
Re=Ul/\nu,
\end{equation}
where $U=\sqrt{\langle \mathbf{u}^2 \rangle}$ is the steady-state root mean square value of the total velocity field and $l$ a characteristic length scale of the flow. Variational methods for optimizing the values of the bounds were later introduced by \cite{doering1994vbe} and \cite{constantin1995vbe} . Since then, these methods have been at the center of the quantitative evaluation of the upper bounds on the dissipation rate in boundary-driven turbulent flows \cite[]{nicodemus1998bfm,kerswell1998uvp}. The first rigorous limits on bulk dissipation for body-force-driven turbulence in a fully periodic domain were derived by \cite{childress2001bdn} and \cite{doering2002edb}. Their result indicates that for the three-dimensional Navier-Stokes equations, the energy dissipation rate $\varepsilon$ satisfies:
\begin{equation}
\label{eq:inequality}
\varepsilon \le c_1 \nu (U^2 / l^2) + c_2 (U^3 / l),
\end{equation} 
where $l$ is the longest characteristic length scale in the body-force function and $c_1$ and $c_2$ are coefficients that depend only on the functional shape (defined more precisely in the next section) of the body-force. Dividing equation \ref{eq:inequality} by $U^3/l$ yields: 
\begin{equation}
\label{eq:inequality_bis}
\beta \le c_1 \frac{1}{Re} + c_2,
\end{equation} 
where $\beta=\varepsilon l/U^3$ is the dimensionless dissipation factor. The behavior of the dissipation factor implied by this relation is in qualitative agreement with theoretical, computational and experimental results for homogeneous isotropic turbulence \cite[]{frisch1995tlk,sreenivasan1984ste,sreenivasan1998ued}. \cite{doering2003edb} derived an explicit upper bound for $\beta$ depending only on the shape of the external forcing at high Reynolds number.  The perspective of determining a value for $\varepsilon$, via $\beta$, directly from the external forcing function presents invaluable advantage since nearly all current turbulent models rely on the prediction of $\varepsilon$ from a solution of a transport equation or semi-empirical models. 

In this short paper, we present the results of a systematic study of the qualitative features of the bounds on the dissipation factor. Few comparisons have been made between the bounds obtained through analytical derivations and the results given by Direct Numerical Simulations (DNS). Moreover, only canonical large scale forcing such as Kolmogorov flow \cite[]{childress2001bdn} or constant shear \cite[]{doering2003edb} has been tested. Using non-trivial forcing functions, we focus, first, on the extent of the $\beta$ dependence on the body-force-shape. Second, we analyze the mean profile dependence on the force profile.  

 \section{Bound on the dissipation factor}
 \subsection{Definitions}
 We are considering a body-force driven incompressible flow on a minimal domain \cite[]{jimenez1991mfu} periodic in three directions. The dynamics of the flow is governed by the Navier-Stokes equations for an incompressible fluid:
 \begin{equation}
 \label{eq:NS}
 \frac{\partial \mathbf{u}}{\partial t}+\left( \mathbf{u} \cdot \boldsymbol{\nabla} \right) \mathbf{u} + \boldsymbol{\nabla}p=\nu \boldsymbol{\nabla^2}\mathbf{u} + \mathbf{f}
 \end{equation}
 \begin{equation}
 \label{eq:incomp}
 \boldsymbol{\nabla} \cdot \mathbf{u}=0,
 \end{equation}
 where $\mathbf{u}$ is the velocity vector, $p$ is the pressure and $\mathbf{f}$ the driving force. The derivation of the bound on the dissipation factor $\beta$ in our case is identical to the one made by \cite{doering2003edb} for the case of a body-forced plane shear flow, except for the boundary conditions.  We therefore limit ourselves to recalling the key points of that derivation and refer to their work for the detailed description of the solution of the variational problem.
 
 The steady force driving the fluid can be written as:
 \begin{equation}
\label{eq:forcing_shape}
\mathbf{f}(\mathbf{x})=F \phi (y/l) \mathbf{e}_x,
\end{equation}
where $l$ is the longest length scale in the forcing function, here the characteristic length of the domain, and $F$ the amplitude. The dimensionless square integrable (or smoother) shape function $\phi \in L^2[0,1]$ satisfies homogeneous Neumann boundary conditions with zero mean:
\begin{equation}
\phi'(0)=\phi'(1), \text{ } \text{ } \text{ } \int_0^1\phi(\eta)d\eta=0.
\end{equation}
The shape function is normalized by:
\begin{equation}
\label{eq:amplitude_forcing}
1=\int_0^1\phi(\eta)^2d\eta.
\end{equation}
defining an unique amplitude $F$ for a given $\mathbf{f}$. For practical purposes, the dimensionless potential $\Phi \in H^1[0,1]$, the space of functions with square-integrable first derivatives, is introduced. $\Phi ' = -\phi$ and $\Phi$ satisfies homogeneous Dirichlet boundary conditions, $\Phi(0)=\Phi(1)$. 

Next, we introduce a mean zero multiplier function $\psi \in H^2[0,1]$ not orthogonal to $\phi$, $\langle \phi \psi \rangle \ne 0$, and satisfying homogeneous Neumann boundary conditions $\psi '(0)  = \psi '(1)$. We define, $\Psi \in H^1 [0,1]$, the derivative of the multiplier function ($\Psi = \psi '$) satisfying homogeneous Dirichlet boundary conditions $\Psi (0) = \Psi (1)$. The inner product of $\phi$ and $\psi$ is equal to the inner product of $\Phi$ and $\Psi$, \textit{i.e.} $ \langle \Phi \Psi \rangle = \langle \phi \psi \rangle \ne 0$.
\subsection{Variational problem and solution}
For a steady state flow, the mean energy injected by the forcing $F \langle \phi u_x \rangle$ should be equal to the total dissipation $\varepsilon$ : 
\begin{equation}
\label{eq:energy_dissipation_rate}
\varepsilon = \nu  \langle \mid \boldsymbol{\nabla} \mathbf{u} \mid^2 \rangle = F \langle \phi u_x \rangle.
\end{equation}

Another expression for the forcing amplitude $F$ can be obtained by projecting the streamwise component of the Navier-Stokes equations onto the multiplier function $\psi$. The inner product of the Navier-Stokes equations with $\psi (y/l) \mathbf{e}_x$ is integrated by parts over the volume. Then the long-time average yields: 
\begin{equation}
\label{eq:innerproduct}
- \left \langle \frac{1}{l} \psi ' u_xu_y \right \rangle = \left \langle \frac{\nu}{l^2} \psi '' u_x \right \rangle +F \langle \phi \psi \rangle.
\end{equation}
Then equation \ref{eq:energy_dissipation_rate}, becomes:
\begin{equation}
\label{eq:epsi}
\varepsilon =- \frac{\langle \phi u_x \rangle \text{ } \left \langle \frac{1}{l} \psi ' u_xu_y + \frac{\nu}{l^2} \psi '' u_x \right \rangle}{\langle \phi \psi \rangle}.
\end{equation}
An expression for the dimensionless dissipation factor $\beta$ is obtained by dividing by $U^3/l$:
\begin{equation}
\label{eq:beta_coarse}
\beta = \frac{\varepsilon l}{U^3} = - \frac{\left \langle \phi \left ( \frac{u_x}{U} \right ) \right \rangle \left \langle \psi ' \left ( \frac{u_x}{U} \right ) \left ( \frac{u_y}{U} \right ) + Re^{-1} \psi '' \left ( \frac{u_x}{U} \right ) \right \rangle}{\langle \phi \psi \rangle}
\end{equation}
Changing the velocity variables to normalized velocities $u \mathbf{e}_x +v \mathbf{e}_y + w \mathbf{e}_z = U^{-1} (u_x \mathbf{e}_x + u_y \mathbf{e}_y + u_z \mathbf{e}_z)$, so that $\langle u^2 + v^2 + w^2 \rangle = 1$, and using the potential $\Phi$ and derivative multiplier $\Psi$, equation \ref{eq:beta_coarse} is recasted as:
\begin{equation}
\label{eq:beta}
\beta = \frac{\langle \Phi ' u \rangle \langle \Psi uv + Re^{-1} \Psi ' u \rangle}{\langle \Phi \Psi \rangle}
\end{equation}
The upper bound $\beta_b$ on the dissipation factor is obtained by maximizing the right-hand side of \ref{eq:beta} over the normalized velocity field, and then minimizing over all multiplier functions $\Psi$:
\begin{equation}
\label{eq:minmaxpb}
\beta \le \underset{\Psi}{\min} \text{ } \underset{\bf{u}}{\max} \text{ } \left[ \frac{\langle \Phi ' u \rangle \langle \Psi uv \rangle}{\langle \Phi \Psi \rangle} + Re^{-1} \frac{\langle \Phi ' u \rangle \langle \Psi ' u \rangle}{\langle \Phi \Psi \rangle} \right]=\beta_b(Re),
\end{equation}
for any solution of the Navier-Stokes equations. $\beta_b$ depends explicitly only on the Reynolds number and on the shape ($\Phi'=-\phi$) of the applied force. 

The evaluation of $\beta_b(Re)$ going beyond the scope of this paper, we refer to \cite{doering2003edb} for that matter and just give their result:
\begin{equation}
\beta_b(Re)=\underset{\Psi}{\min} \text{ } \frac{\langle \phi ^2 \rangle^{1/2}}{\langle \Phi \Psi \rangle} \left[ \frac{1}{\sqrt{27}} \text{ } \underset{y \in \left[ 0,1 \right]}{\sup} \mid \Psi (y) \mid + Re^{-1} \langle \Psi'^2 \rangle^{1/2} \right].
\end{equation}
Finally, these authors solve exactly the extremization problem for the optimal $\Psi$ in the asymptotic case $Re \rightarrow \infty$:
\begin{equation}
\underset{Re \rightarrow \infty}{\lim \text{ } \sup} \text{ } \beta_b (Re) \le \beta_b(\infty)=\frac{1}{\sqrt{27}}\text{ } \frac{\sqrt{\langle \phi^2 \rangle}}{\langle \mid\textbf{$\Phi$}\mid \rangle}
\end{equation}
The shape function being normalized (\textit{i.e.} $\langle \phi^2 \rangle=1$), we can simplify this bound a little further:
\begin{equation}
\label{eq:beta_inf}
\beta_b(\infty)= \frac{1}{\sqrt{27} \langle \mid\textbf{$\Phi$}\mid \rangle}
\end{equation}
 
 \section{Force-shape dependence of the dissipation factor}
A non-trivial body-force-shape is tested based on the Kolmogorov forcing. A second wave number is added with different amplitudes $A_k$ to the wave number $k=1$ used in the traditional Kolmogorov forcing:
\begin{equation}
\label{eq:2modes}
f(\mathbf{x})=\left[ \sin (2\pi\eta)+A_k \sin (2\pi k\eta) \right] \mathbf{e}_x=F\phi(\eta)\text{ }\mathbf{e}_x, \text{ } \text{ } \text{ } \eta\in \left[ 0,1 \right], \text{ } k \ge 1
\end{equation}
where,
\begin{equation}
F=\sqrt{\frac{1+A_k^2}{2}} \text{ } \text{ } \text{and} \text{ } \text{ } \phi(\eta)=\sqrt{\frac{2}{1+A_k^2}}\text{ } \left[\sin (2\pi\eta)+A_k \sin (2\pi k\eta) \right],
\end{equation}
with $A_1=0$ and $A_k \in \Re$ for $k \ge 2$. In the following, the term representing the traditional Kolmogorov forcing in equation \ref{eq:2modes} will often be referred as ``primary term" and the term of higher wave number as ``secondary term". 

We can now use \ref{eq:beta_inf} in order to evaluate the bound on $\beta$ in the asymptotic case $Re \rightarrow \infty$ for the non-trivial body-force-shape define above:
\begin{equation}
\label{eq:case2}
\beta_b(\infty)=\frac{1}{\sqrt{27}}\text{ } \frac{1}{\int_0^1\mid\frac{1}{2\pi} \sqrt{\frac{2}{1+A_k^2}}\text{ } (\cos (2\pi\eta) + \frac{A_k}{k}\text{ } \cos (2\pi k\eta))\mid d\eta}.
\end{equation}
The integral on the denominator of the right hand side of equation \ref{eq:case2} can be rather painful to solve by hand. First, we consider the simple cases.

The case of the classical Kolmogorov forcing, $k=1$ ($A_1=0$), is straightforward. Equation \ref{eq:case2} becomes:
\begin{equation}
\label{eq:betakolmo}
\beta_b(\infty)=\frac{1}{\sqrt{27}}\text{ } \frac{1}{\int_0^1\mid\frac{1}{2\pi}\sqrt{2}\text{ } \cos (2\pi\eta)\mid d\eta}=\frac{\pi^2}{\sqrt{54}}.
\end{equation}
This result indicates that for large enough Reynolds numbers, the dissipation factor for the Kolmogorov forcing is bound from above by a constant value equal to $\pi^2/\sqrt{54}$. This value holds whatever is the amplitude of the Kolmogorov forcing.

Another trivial case appears when the secondary term becomes dominant (\textit{i.e.} $A_k \gg 1$). The contribution of the primary term to the forcing function can therefore be neglected and the forcing becomes equivalent to a Kolmogorov forcing at a wave number greater than one:
\begin{equation}
\label{eq:betaint}
\beta_b(\infty)=\frac{1}{\sqrt{27}}\text{ } \frac{1}{\frac{1}{2\pi}\text{ } \sqrt{\frac{2}{A_k^2}}\text{ }\int_0^1\mid \frac{A_k}{k}\text{ } \cos (2\pi k\eta)\mid d\eta}=\frac{k\pi^2}{\sqrt{54}}.
\end{equation}
The upper bound on $\beta$ in the asymptotic case $Re \rightarrow \infty$ is linearly proportional to the wave number of the largely dominant term in the forcing function. It is not surprising that when a single mode is largely dominant in the force shape functional, the bound on the dissipation factor $\beta$ is related to this particular mode. However, the linear increase of $\beta_b(\infty)$ as a function of the dominant wave number is not intuitive. This result implies that for a trivial force-shape, in other words a force-shape dictated by a single wave number, we could predict precisely the dissipation factor in the ideal case of an infinite Reynolds number. Each wave number forced alone is bound by a characteristic $\beta_b$ when $Re \rightarrow \infty$.
\begin{figure}
\begin{center}
\centerline{\includegraphics[height=6.7cm,width=6.7cm,angle=0]{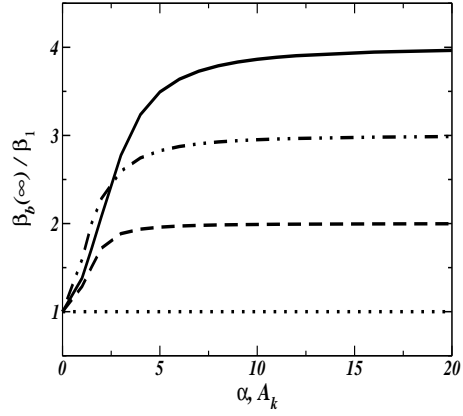}}
\end{center}
\caption{\label{fig:boundth} Solution of equation \ref{eq:case2} normalized by $\beta_{1} \equiv \beta_b(\infty)_{\mid f(\mathbf{x})=\alpha \sin(2\pi\eta), \alpha \in \Re}$ plotted as a function of the amplitude of the secondary mode forced, $A_k$ (or $\alpha$ in the case of the classic Kolmogorov forcing). Dotted line: $f_x=\alpha \sin (2\pi \eta), \alpha \in \Re$; Dashed line: $f_x=\sin (2\pi\eta)+A_2\text{ }\sin (2\times 2\pi\eta)$; Dash-dot-dot line: $f_x=\sin (2\pi\eta)+A_3\text{ }\sin (3\times 2\pi\eta)$; solid line: $f_x=\sin (2\pi\eta)+A_4\text{ }\sin (4\times 2\pi\eta)$.}
\end{figure}

To further understand the force-shape dependence of the bound on the dissipation factor behavior, a complete resolution of equation \ref{eq:case2} is required. Figure \ref{fig:boundth} shows the evolution of the bound on the dissipation factor in the asymptotic case, $\beta_b(\infty)$, for a non-trivial force-shape normalized by the value of $\beta_b(\infty)$ obtained for a classic Kolmogorov forcing (\textit{i.e.}, $\pi^2/\sqrt{54}$), as a function of the amplitude of the secondary term. In addition to the properties exposed by equations \ref{eq:betakolmo} and \ref{eq:betaint}, this figure also shows that $\beta_b(\infty)$ is very sensitive to the change in shape of the forcing function. Indeed, as soon as a secondary term is added to the original Kolmogorov forcing, $\beta_b(\infty)$ increases in a quasi-linear fashion as a function of the amplitude of the secondary term. This quasi-linear increase lasts until the rate of variation of $\beta_b(\infty)$ is greater than about $80\%$. The increase of $\beta_b(\infty)$ then drastically slows down and plateaus. One can consider that the secondary term is largely dominant beyond this point. We can further anticipate that the characteristic features of the secondary term are the most obvious in the force profile, the contribution of the primary term being barely visible. We observe that $\beta_b(\infty)$ increases the fastest when $A_3\sin (3\times 2\pi\eta)$ is added to the classic Kolmogorov forcing, for small values of $A_3$. We verify however that this feature is isolated and not linked to the odd wave numbers.

\section{Comparison to Direct Numerical Simulations}
\begin{figure}
  \centering
\centerline{\includegraphics[height=6.7cm,width=6.7cm,angle=0]{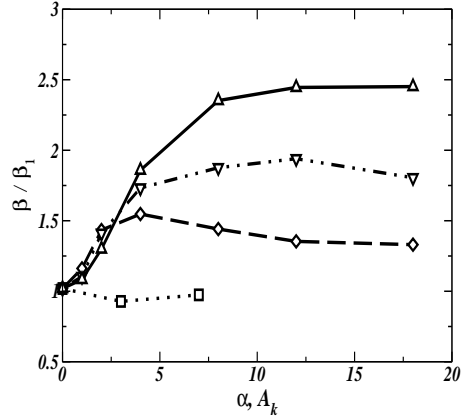}}
\caption{Dissipation factor, $\beta$, obtained by DNS and normalized by $\beta_{1} \equiv \beta_{\mid f(\mathbf{x})=\alpha \sin(2\pi\eta), \alpha \in \Re}$ plotted as a function of the amplitude of the secondary mode forced, $A_k$ (or $\alpha$ in the case of the classic Kolmogorov forcing). Squares, dotted line: $f=\alpha \text{ }\sin (2\pi\eta)$, $\alpha =1, 3, 7$; diamonds, dashed line: $f=\sin (2\pi\eta)+A_2\text{ }\sin (2\times 2\pi\eta)$; down triangles, dash-dot-dot line: $f=\sin (2\pi\eta)+A_3\text{ }\sin (3\times \pi\eta)$; up triangle, solid line: $f=\sin (2\pi\eta)+A_4\text{ }\sin (4\times 2\pi\eta)$.}
  \label{fig:beta_amplitude}
\end{figure}
Our Direct Numerical Simulations are performed using a fully de-aliased spectral code. The time stepping is based on a third-order Runge-Kutta algorithm for the non-linear and forcing terms. The viscous term is integrated through an analytic factor. The stability is ensured by a CFL condition. The flow is solved in a $2\pi$ in length periodic cubic box with $128\times128\times128$ grid points. The viscosity is set to $\nu=0.015625$, large enough to ensure full spectrum resolution of the turbulent flow in our domain.

Figure \ref{fig:beta_amplitude} shows the dissipation factor $\beta$ normalized by the average value of $\beta$ in the case of the classic Kolmogorov forcing at different amplitudes as a function of the amplitude of the secondary term. Each symbol represents a simulation with a different force shape following the definition of equation \ref{eq:2modes}. The symbols designing a particular secondary wave numder $k$ are linked by straight lines using the same nomenclature as in figure \ref{fig:boundth} in order to ease comparisons between the behaviors of $\beta_b(\infty)$ and $\beta$. We verify that all the simulations satisfy the convergence criterion for the Kolmogorov flow defined by \cite{sarris2007bsd}. The simulations with secondary term amplitudes of $A_k=18$ verify $k_{max}\eta > 1.2$, where $k_{max}$ is the largest wave number of the simulation and $\eta$ the Kolmogorov length scale, slightly under the commonly used flow resolution criterion $k_{max}\eta > 1.5$. 

On figure \ref{fig:beta_amplitude},  we recognize the same patterns displayed on figure \ref{fig:boundth}. 
First, $\beta$ has a nearly constant value when only the wave number $k=1$ is forced. In this particular case, a limited number of simulations was possible as for an amplitude of $7$ for the forcing, the limit of resolution of our flow domain was already achieved. As regard to the force shape, we can anticipate that the value of $\beta$ will remain about constant for any amplitude.

Second, in the case of a non-trivial force-shape, $\beta$ increases in a nearly linear fashion as a function of $A_k$ until the secondary term becomes dominant ($A_k$ sufficiently large). Then, the increase in $\beta$ slows down and plateaus. The analytic bound on the dissipation factor in the asymptotic case of an infinite Reynolds number predicts qualitatively the behavior of the dissipation factor at our low Reynolds number Direct Numerical Simulations. Indeed, our simulations have Taylor-scale Reynolds number, $R_{\lambda}$, varying between about $50$ and $100$. The observations made from the predicted bound on the dissipation factor when $Re \rightarrow \infty$ apply at low to moderate $Re$: \textit{i}) $\beta$ is very sensitive to changes in the shape of the forcing function, \textit{ii}) $\beta$ saturates at a characteristic level for a given wave number and \textit{iii}) it seems that the level of saturation of $\beta$ is linearly proportional to the wave number.

The symmetry in the qualitative results obtain from figure \ref{fig:boundth} and \ref{fig:beta_amplitude} proves that $\beta$ is not only sensitive to the force-shape as observed above, but $\beta$ strongly depends on it. First, when a single wave number is forced, a variation on the amplitude of the forcing function alone implies no variation on the shape (\textit{i.e.}, the general appearance of the profile) but a linear variation of the slopes of  the force-shape as a function of the amplitude. The variations of amplitude affect the total kinetic energy of the flow, so it affects $U$ (as defined in this paper, $U \propto E^{1/2}$, where $E$ is the total kinetic energy). The variations of the slopes affect the kinetic energy dissipation $\varepsilon$. In addition, according to Kolmogorov's third law, $\varepsilon \propto E^{3/2} \equiv \varepsilon \propto U^3$, for sufficiently high Reynolds numbers. Therefore, having $\beta$ (about) constant is directly related to the steadiness of the force-shape in this particular case. The same reasoning applies when a given wave number is largely dominant in the force-shape. Second, when a secondary term with a small enough amplitude is added to the forcing on a single wave number, the amplitude of the new forcing function varies barely whereas the force-shape becomes significantly different (see case $A_k=1$ on figure \ref{fig:force}  for example). The total kinetic energy of the flow changes barely and the kinetic energy dissipation changes significantly. As a result, we observe a significant variation in $\beta$ ($\beta \propto \varepsilon/E^{3/2}$). We can anticipate the same type of influence on the dissipation factor for secondary terms with large amplitude, but not large enough for the secondary term to be largely dominant.

As already observed, despite qualitative agreement, the numerical values for $\beta$ are about a factor $5$ below the predicted upper bound (not visible here because of the normalization of $\beta$ in the figures) \cite[]{doering2003edb}. Also figures \ref{fig:boundth} and \ref{fig:beta_amplitude} clearly show that the theory does not predict correctly the magnitude of the increase in $\beta$ when forcing an additional wave number. The slopes in the quasi-linear part of the evolution of $\beta$ are greater in figure \ref{fig:boundth} than in figure \ref{fig:beta_amplitude}.  
   
\section{Mean velocity profile dependence on the force profile}
\begin{figure}
  \centering
  \subfloat[]{\label{fig:S_2_1}\includegraphics[height=6.7cm,width=6.7cm,angle=0]{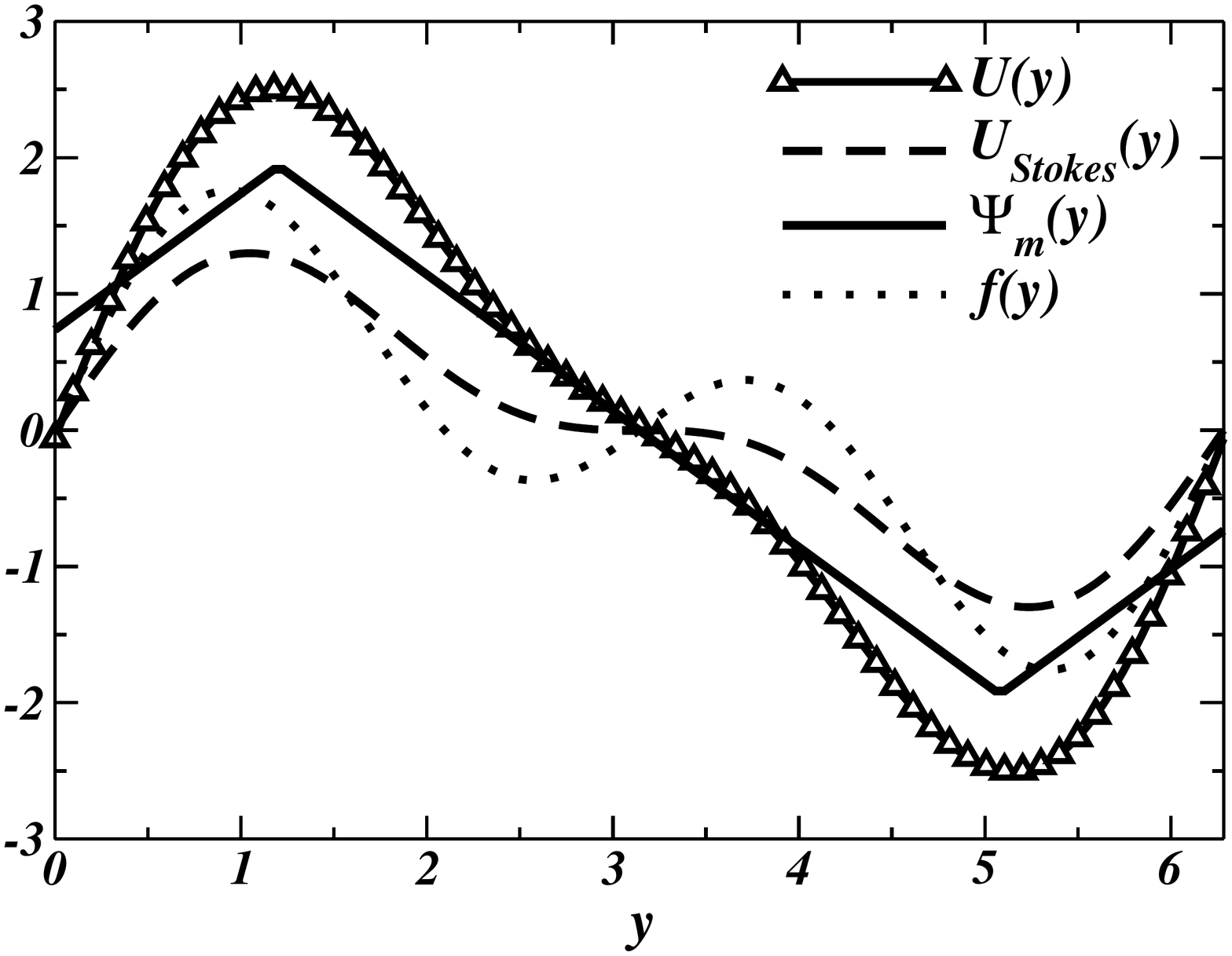}}                
  \subfloat[]{\label{fig:S_3_1}\includegraphics[height=6.7cm,width=6.7cm,angle=0]{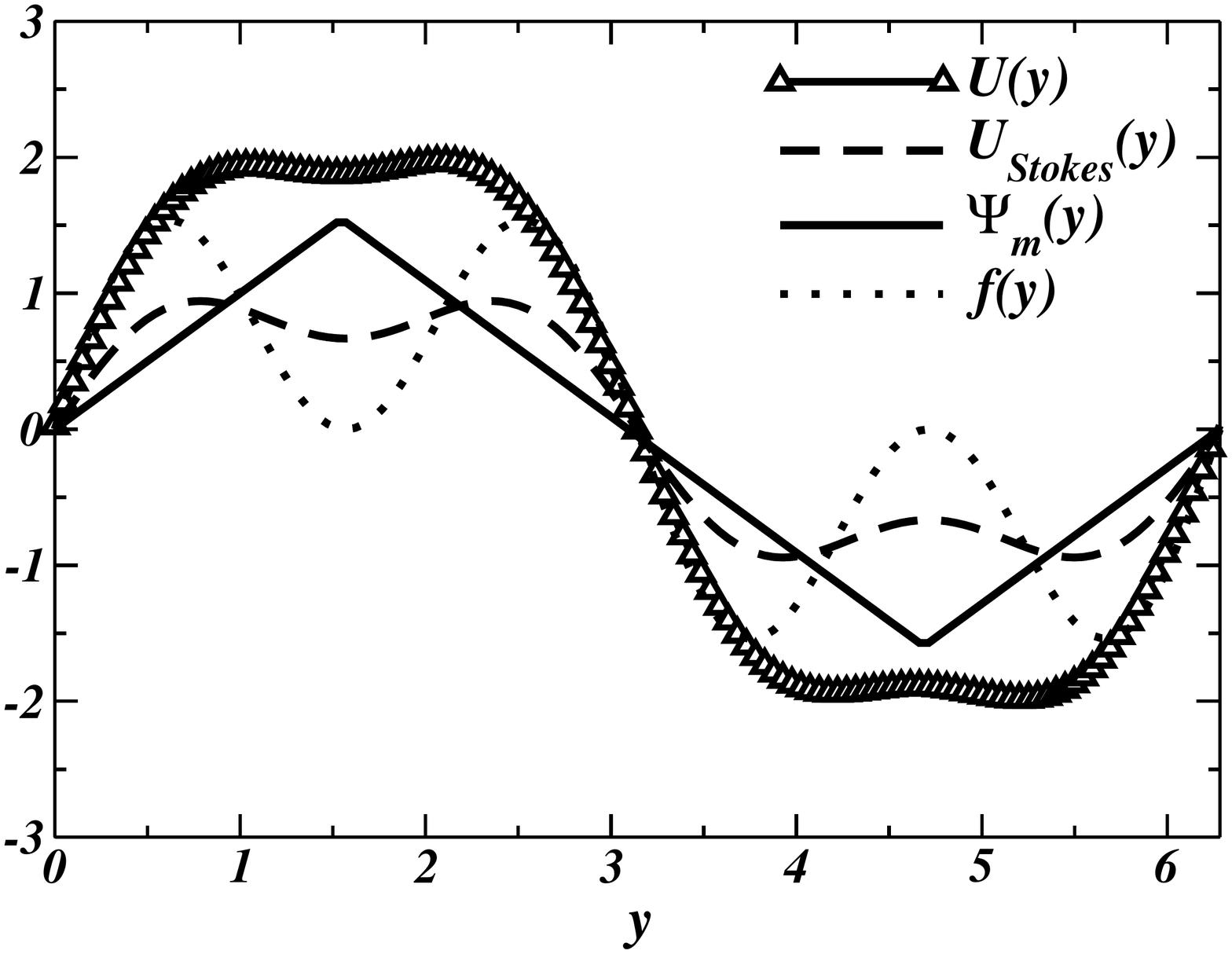}}                
  \subfloat[]{\label{fig:S_4_1}\includegraphics[height=6.7cm,width=6.7cm,angle=0]{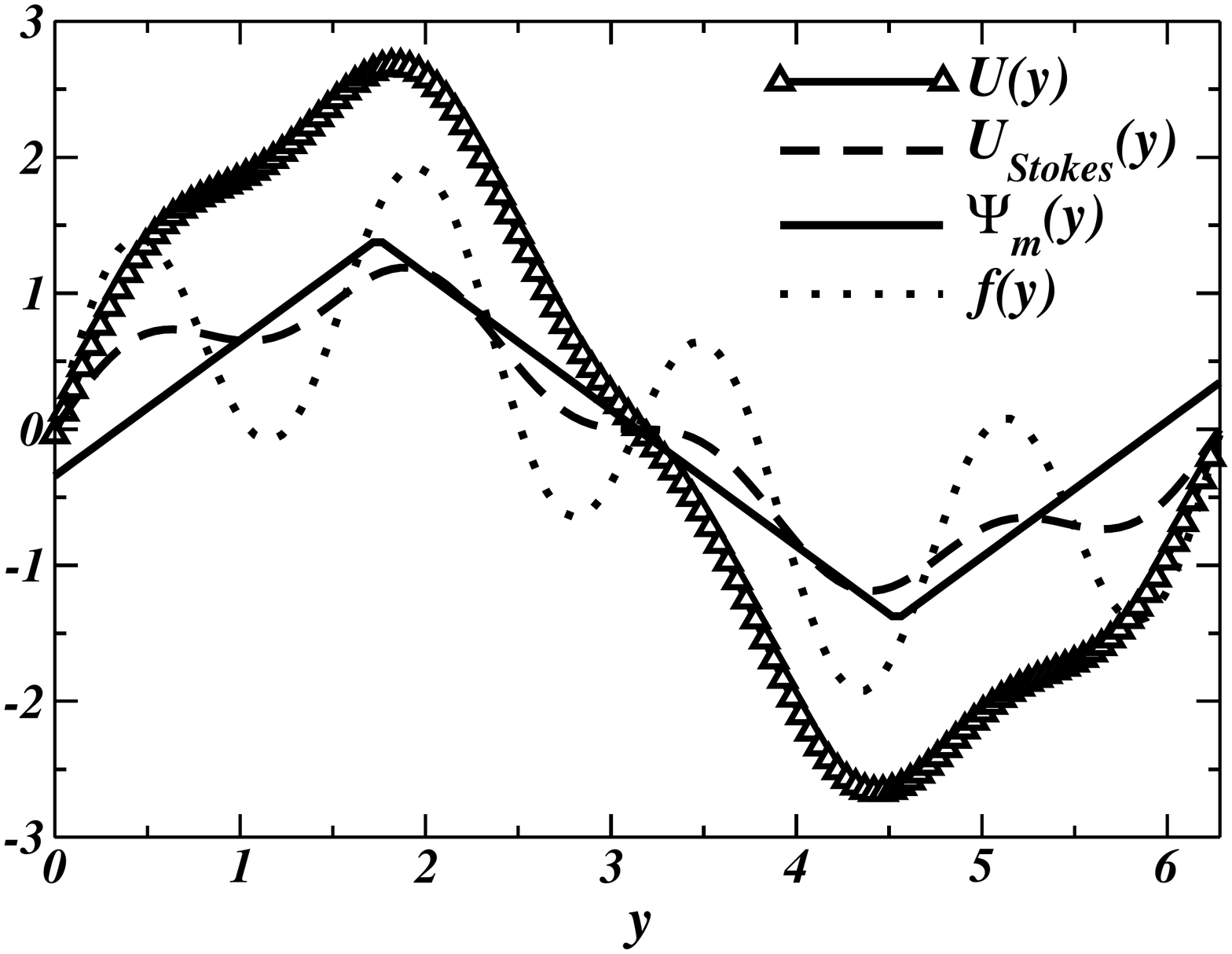}}                  
\caption{Mean velocity profile $U(y)$ (solid line and triangles), Stokes flow profile $U_{Stokes}(y)$ (dashed line), optimal multiplier profile $\Psi_m(y)$ (solid line) and corresponding forcing profile $f(y)$ (dotted line). \textit{a}) $f(y)=sin(y)+sin(2y)$, \textit{b}) $f(y)=sin(y)+sin(3y)$, \textit{c}) $f(y)=sin(y)+sin(4y)$.}
  \label{fig:stokes}
\end{figure}
\begin{figure}
  \centering
  \subfloat[]{\label{fig:Multi}\includegraphics[height=6.7cm,width=6.7cm,angle=0]{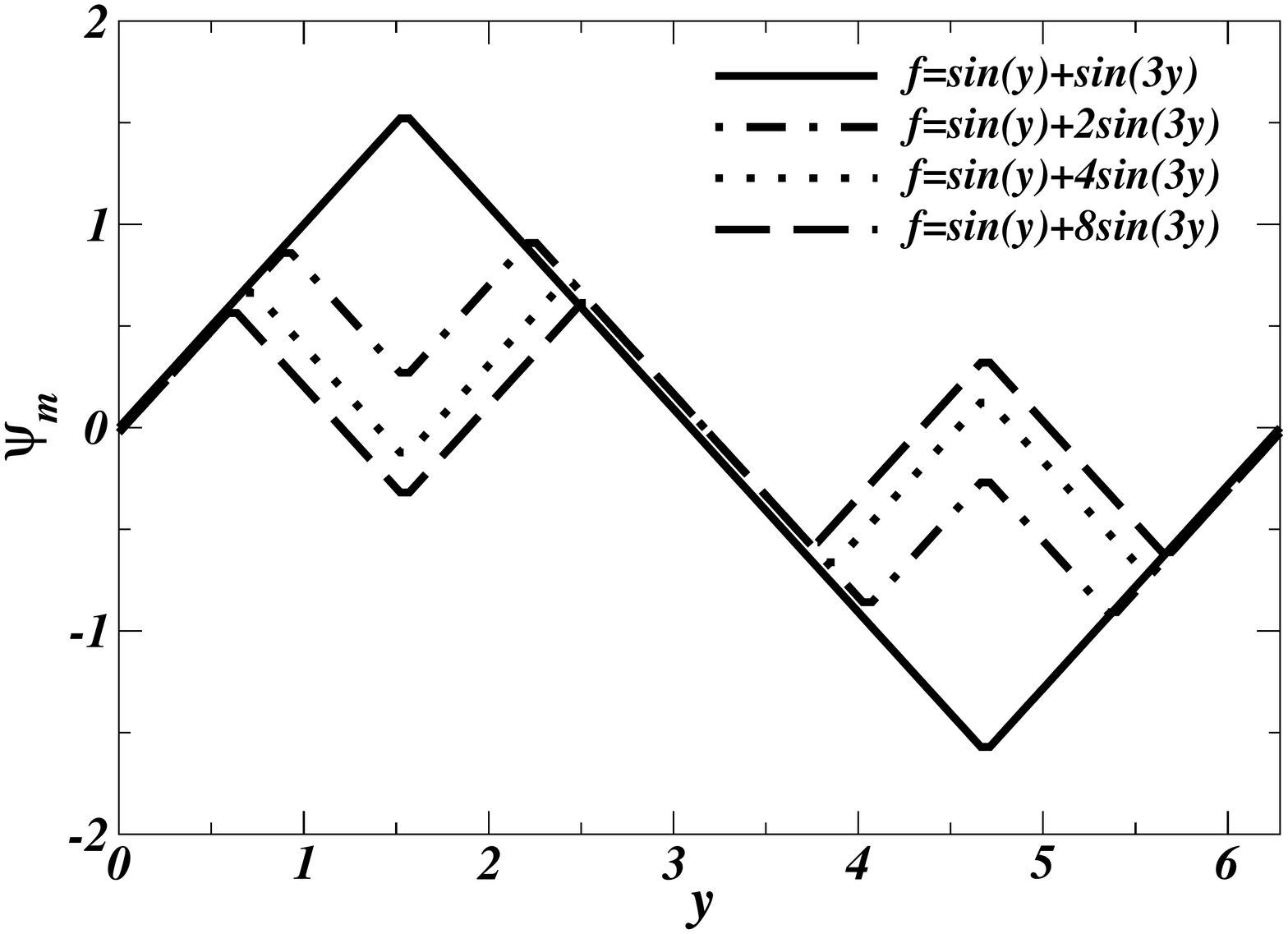}}                
  \subfloat[]{\label{fig:Mprof}\includegraphics[height=6.7cm,width=6.7cm,angle=0]{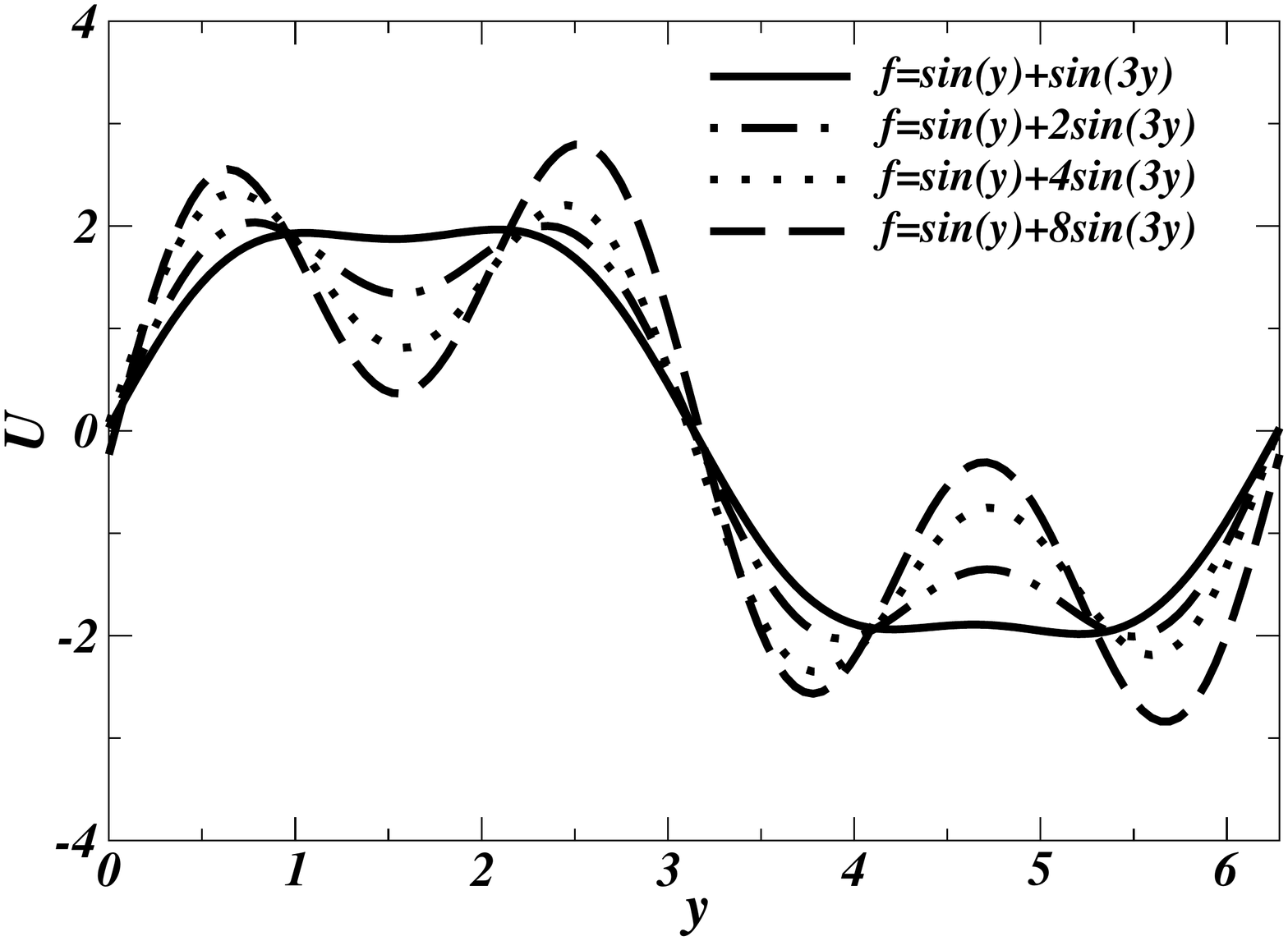}} 
  \subfloat[]{\label{fig:force}\includegraphics[height=6.7cm,width=6.7cm,angle=0]{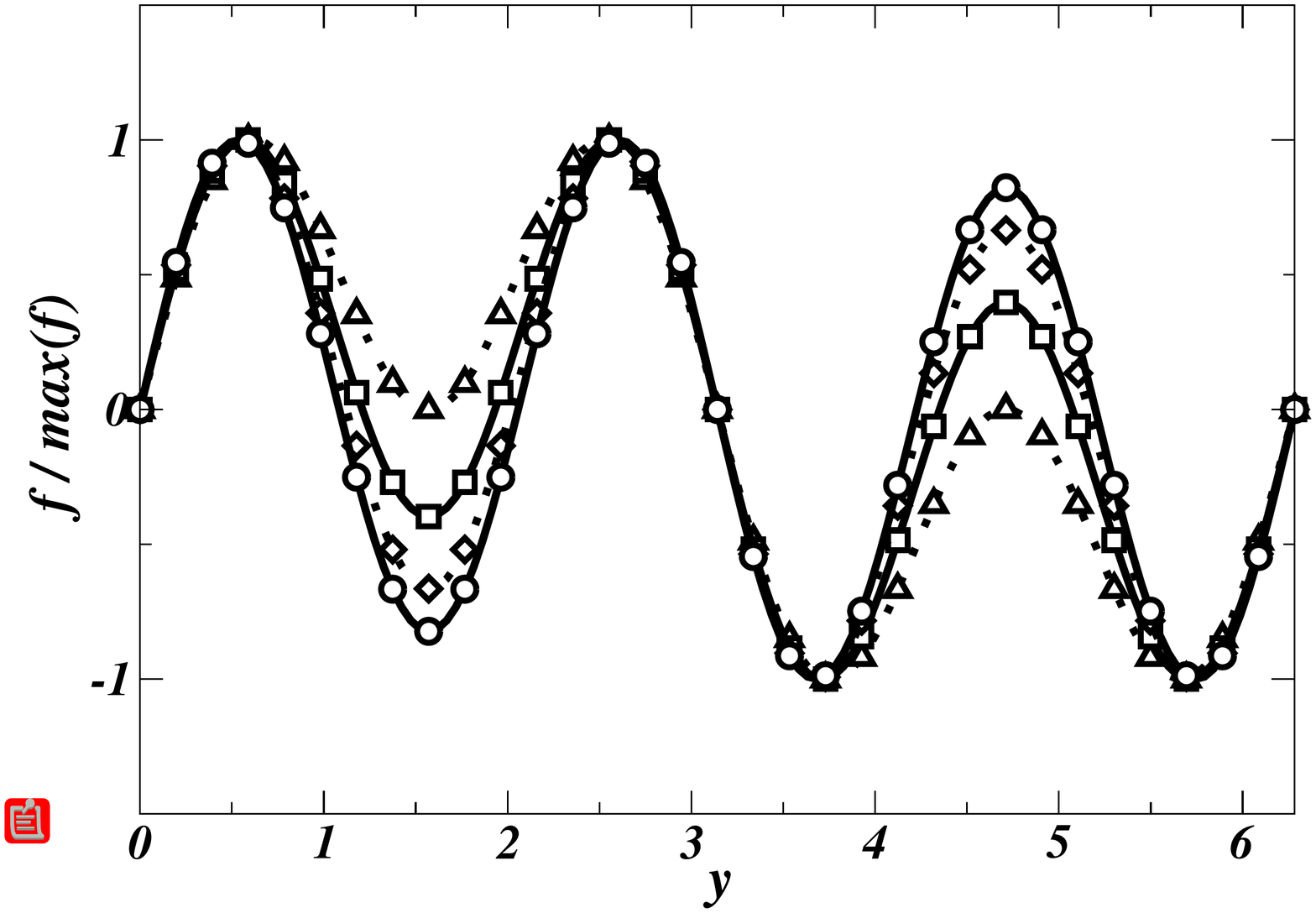}}                
 \caption{\textit{a}) Optimal multiplier profile $\psi_m(y)$ for the force profile $f(y)=sin(y)+A_ksin(3y)$, $A_k=1, 2, 4, 8$. \textit{b}) Mean velocity profile $U(y)$ for the force profile $f(y)=sin(y)+A_ksin(3y)$, $A_k=1, 2, 4, 8$. \textit{c}) Force profile $f(y)=sin(y)+A_ksin(3y)$, $A_k=1, 2, 4, 8$, normalized by $\max (f)$, its maximum value. Triangles on dotted line: $A_k=1$; squares on solid line: $A_k=2$; diamonds on dotted line: $A_k=4$; circles on solid line: $A_k=8$.}
  \label{fig:PM}
\end{figure}
Figure \ref{fig:stokes} shows the mean velocity profile $U(y)$, the Stokes flow profile $U_{stokes}(y)$, the optimal multiplier profile $\psi_m(y)$ and the corresponding forcing function $f(y)=sin(y)+sin(ky)$, where $k=2,3,4$ respectively in $a)$, $b)$ and $c)$. The Stokes flow is the flow associated with a lower bound on the dissipation factor $\beta$.
The Stokes flow profile is simply obtained by integrating twice the forcing function:
\begin{equation}
U_{Stokes}(y)=\int_0^y \left( \int_0^{y'} F\phi(y'')dy''\right) dy'+C,
\end{equation} 
where $C$ is a constant always equal to $0$ in our configuration. The optimal multiplier $\psi_m$ in the case of infinite Reynolds numbers is given by:
\begin{equation}
\psi_m(y)=\int_0^y sgn\left( \int_0^{y'}\phi(y'')dy''\right)dy'+C,
\end{equation}
where $C$ is a constant determined by using the zero mean condition on $\phi$. We made here the hypothesis that the optimal multiplier at finite Reynolds numbers is identical to the infinite Reynolds numbers optimal multiplier. 

In the cases presented on figure \ref{fig:stokes}, the low wave number term in the forcing function is the dominant term. Contrary to the DNS presented by \cite{doering2003edb}, our domain is periodic in all three directions. There are no free-slip boundary conditions to constrain the velocity profiles. Thus, the key feature of these velocity profiles is that they display all the characteristic elements of the force profile. These elements appear to various extent whether we have a Stokes flow or a fully developed turbulent flow. For the velocity profiles shown in these figures, the dominant shape is clearly a sine function because of the small contribution of the secondary term. But, we see by looking at the force profiles that a well defined change in slope induces a change in slope for the velocity profiles. These changes are more obvious in Stokes profiles (figure \ref{fig:stokes}) than in the turbulent flow mean velocity profiles. Indeed, in the turbulent case, a deviation in the force profile as seen on figure \ref{fig:S_2_1} around $y=\pi$ creates too little of a shear on a too little sub-domain compared to the combined effect of the shear on both sides of this sub-domain. Functions of high wave number provide less energetic contribution to the turbulent flow than functions of low wave number. Therefore, the weight on the secondary term of the forcing function must be greater than the weight on the primary term in order to have at least an equivalent contribution to the turbulent flow. Hence the minute contribution to the Stokes and turbulent mean velocity profiles of the secondary term in figure \ref{fig:stokes}. The optimal multiplier profiles relate to the dominant term of the forcing functions. They appear to indicate only significant contributions to the force profile, here the primary term. 

Figure \ref{fig:PM} allows a broader perspective on behavior of the optimal multiplier $\psi_m$. It shows the optimal multiplier profile, the mean velocity profile and the normalized force profile as the amplitude on the secondary mode is increased. As mentioned above, $\psi_m$ appears to be less sensitive than the mean velocity profile as regard to the forcing profile. Indeed, we verify that for the cases $f(y)=sin(y)$ and $f(y)=sin(y)+sin(3y)$, $\psi_m$ displays the same profile. For larger amplitudes ($A_k \ge 2$) on the secondary term, $\psi_m$ displays a shape characteristic to the wave numbers composing the force-shape. As the amplitude on the secondary term increases, the amplitude of $\psi_m$ decreases (figure \ref{fig:Multi}). The decrease in the amplitude of $\psi_m$ is slower as the amplitude on the secondary term gets larger. We can relate the evolution of $\psi_m$ to the evolution of the slopes of the features  characteristic to the secondary term in the force-shape profile. In figure \ref{fig:force}, the local minimum at $\pi/4$ and local maximum at $3\pi/4$ are caused by the addition of the secondary term. We observe that the variation of the slopes around these points is smaller as the amplitude on the secondary term increases. We verify that this variation of slope becomes constant as the secondary term becomes largely dominant (contribution of the primary term negligible). Therefore, amplitude of the optimal multiplier remains constant when the secondary term is largely dominant. Finally, figure \ref{fig:Mprof} shows that the relation between $\psi_m$ and the mean velocity profile is unclear. For $f(y)=sin(y)+sin(3y)$, the optimal multiplier shows no indication of the influence of the secondary term whereas the mean velocity profile does. 

\section{Conclusions}
To summarize, low Reynolds number Direct Numerical Simulations confirmed qualitative results predicted by the mathematical analysis at infinite Reynolds numbers. This confirms the dependence of the dissipation factor on the force-shape as predicted by the solution of the optimization problem. We can discern two major tendencies: \text{i) } when the forcing function is mainly shaped by a single wave number, the dissipation factor is unique, its value being seemingly proportional to the dominant wave number. \textit{ii}) $\beta$ increases in a quasi-linear fashion when an additional term is added to the forcing function until this additional term becomes dominant.

Also, we observed than the mean velocity profiles display all the characteristic features of the forcing function in a 3D-periodic domain. The extent of these features are modulated as a function of the contribution of each component on the forcing function. As a consequence, in 3D-periodic domains, it should be possible to manipulate the force-shape in order to obtain a given mean velocity profile. The optimal multiplier evolution is tied to the force-shape, but its relation to the mean velocity profile remains unclear.

In this short paper, we showed that in unbound turbulence, the force-shape plays a major role in the energy dissipation rate. The optimization problem solved by \cite{doering2003edb} captures this dependence qualitatively for any forcing function. We demonstrated that their high Reynolds numbers prediction remains valid for moderate Reynolds numbers. Improvements in the quantitative prediction of the dissipation factor remain necessary in order to be able to use the dissipation in a turbulence model.\\
\\
The computational resources provided by the Vermont Advanced Computing Center, which is supported by NASA (Grant No. NNX 06AC88G), are gratefully acknowledged. We are also grateful to Professors Carati and Knaepen and all the Turbo team for providing the code. 

\bibliographystyle{jfm}
\bibliography{bib_VKF_Rollin}

\begin{thebibliography}{18}
\expandafter\ifx\csname natexlab\endcsname\relax\def\natexlab#1{#1}\fi

\bibitem[Childress {\em et~al.\/}(2001)Childress, Kerswell \&
  Gilbert]{childress2001bdn}
{\sc Childress, S., Kerswell, RR \& Gilbert, AD} 2001 {Bounds on dissipation
  for Navier--Stokes flow with Kolmogorov forcing}. {\em Physica D: Nonlinear
  Phenomena\/} {\bf 158}~(1-4), 105--128.

\bibitem[Constantin \& Doering(1995)]{constantin1995vbe}
{\sc Constantin, P. \& Doering, C.R.} 1995 {Variational bounds on energy
  dissipation in incompressible flows. II. Channel flow}. {\em Physical Review
  E\/} {\bf 51}~(4), 3192--3198.

\bibitem[Doering \& Constantin(1992)]{doering1992eds}
{\sc Doering, C.R. \& Constantin, P.} 1992 {Energy dissipation in shear driven
  turbulence}. {\em Physical Review Letters\/} {\bf 69}~(11), 1648--1651.

\bibitem[Doering \& Constantin(1994)]{doering1994vbe}
{\sc Doering, C.R. \& Constantin, P.} 1994 {Variational bounds on energy
  dissipation in incompressible flows: Shear flow}. {\em Physical Review E\/}
  {\bf 49}~(5), 4087--4099.

\bibitem[Doering {\em et~al.\/}(2003)Doering, Eckhardt \&
  Schumacher]{doering2003edb}
{\sc Doering, C.R., Eckhardt, B. \& Schumacher, J.} 2003 {Energy dissipation in
  body-forced plane shear flow}. {\em Journal of Fluid Mechanics\/} {\bf 494},
  275--284.

\bibitem[Doering \& Foias(2002)]{doering2002edb}
{\sc Doering, C.R. \& Foias, C.} 2002 {Energy dissipation in body-forced
  turbulence}. {\em Journal of Fluid Mechanics\/} {\bf 467}, 289--306.

\bibitem[Frisch(1995)]{frisch1995tlk}
{\sc Frisch, U.} 1995 {\em {Turbulence: The Legacy of AN Kolmogorov}\/}.
  Cambridge University Press.

\bibitem[Jim{\'e}nez \& Moin(1991)]{jimenez1991mfu}
{\sc Jim{\'e}nez, J. \& Moin, P.} 1991 {The minimal flow unit in near-wall
  turbulence}. {\em Journal of Fluid Mechanics\/} {\bf 225}, 213--240.

\bibitem[Kerswell(1998)]{kerswell1998uvp}
{\sc Kerswell, RR} 1998 {Unification of variational principles for turbulent
  shear flows: the background method of Doering-Constantin and the
  mean-fluctuation formulation of Howard-Busse}. {\em Physica D: Nonlinear
  Phenomena\/} {\bf 121}~(1-2), 175--192.

\bibitem[Kolmogorov(1941)]{kolmogorov1941lst}
{\sc Kolmogorov, A.} 1941 {The Local Structure of Turbulence in Incompressible
  Viscous Fluid for Very Large Reynolds' Numbers}. {\em Dokl. Akad. Nauk
  SSSR\/} {\bf 30}, 301--305.

\bibitem[Marchioro(1994)]{marchioro1994red}
{\sc Marchioro, C.} 1994 {Remark on the energy dissipation in shear driven
  turbulence}. {\em Physica D\/} {\bf 74}~(3-4), 395--398.

\bibitem[Nicodemus {\em et~al.\/}(1998)Nicodemus, Grossmann \&
  Holthaus]{nicodemus1998bfm}
{\sc Nicodemus, R., Grossmann, S. \& Holthaus, M.} 1998 {The background flow
  method. Part 1. Constructive approach to bounds on energy dissipation}. {\em
  Journal of Fluid Mechanics\/} {\bf 363}, 281--300.

\bibitem[Richardson(1922)]{richardson1922wpn}
{\sc Richardson, L.F.} 1922 {\em {Weather Prediction by Numerical Process}\/}.
  University Press.

\bibitem[Sarris {\em et~al.\/}(2007)Sarris, Jeanmart, Carati \&
  Winckelmans]{sarris2007bsd}
{\sc Sarris, I.E., Jeanmart, H., Carati, D. \& Winckelmans, G.} 2007 {Box-size
  dependence and breaking of translational invariance in the velocity
  statistics computed from three-dimensional turbulent Kolmogorov flows}. {\em
  Physics of Fluids\/} {\bf 19}, 095101.

\bibitem[Sreenivasan(1984)]{sreenivasan1984ste}
{\sc Sreenivasan, KR} 1984 {On the scaling of the turbulence energy dissipation
  rate}. {\em Physics of Fluids\/} {\bf 27}, 1048.

\bibitem[Sreenivasan(1998)]{sreenivasan1998ued}
{\sc Sreenivasan, K.R.} 1998 {An update on the energy dissipation rate in
  isotropic turbulence}. {\em Physics of Fluids\/} {\bf 10}, 528.

\bibitem[Taylor(1938)]{taylor1938st}
{\sc Taylor, GI} 1938 {The Spectrum of Turbulence}. {\em Proceedings of the
  Royal Society of London. Series A, Mathematical and Physical Sciences\/} {\bf
  164}~(919), 476--490.

\bibitem[Wang(1997)]{wang1997tae}
{\sc Wang, X.} 1997 {Time averaged energy dissipation rate for shear driven
  flows in R n}. {\em Physica D: Nonlinear Phenomena\/} {\bf 99}~(4), 555--563.

\end{thebibliography}

\end{document}